\documentclass{ws-mplb}

\begin{document}

\markboth{I. Hasegawa and H. Shima}{CONTINUOUS TRANSITION OF 
DEFECT CONFIGURATION IN A DEFORMED LIQUID CRYSTAL FILM}

%
\catchline{}{}{}{}{}
%

\title{CONTINUOUS TRANSITION OF DEFECT CONFIGURATION 
IN A DEFORMED LIQUID CRYSTAL FILM}

\author{Isaku Hasegawa}

\address{Division of Applied Physics, Graduate School of Engineering, 
Hokkaido University, Kita13, Nishi8, Kita-ku, Sapporo, Hokkaido 
060-8628, Japan\\
isaku@eng.hokudai.ac.jp}

\author{Hiroyuki Shima}

\address{Division of Applied Physics, Faculty of Engineering, 
Hokkaido University, Kita13, Nishi8, Kita-ku, Sapporo, Hokkaido 
060-8628, Japan\\
Department of Applied Mathematics 3, LaC\`aN, Universitat 
Polit\`ecnica de Catalunya, Barcelona 08034, Spain\\
shima@eng.hokudai.ac.jp}

\maketitle

\begin{history}
\received{(Day Month Year)}
\revised{(Day Month Year)}
\end{history}

\begin{abstract}
We investigate energetically favorable configurations of point 
disclinations in nematic films having a bump geometry.
Gradual expansion in the bump width $\Delta$ gives rise to a sudden shift 
in the stable position of the disclinations from the top to the skirt of the 
bump.
The positional shift observed across a threshold $\Delta^{\rm th}$ obeys a 
power law function of $\left| \Delta - \Delta^{\rm th} \right|$, 
indicating a new class of continuous phase transition that governs the defect 
configuration in curved nematic films.
\end{abstract}


\section{Introduction}
Optimal control of molecular alignment is crucial for applications of 
liquid crystal films.
Thus far, many sophisticated techniques  have been developed for obtaining 
homogeneous alignment in a desired direction: mechanical rubbing treatment 
\cite{berreman,fukuda1,fukuda2}, 
surface-polarization effect \cite{barbero1,barbero2}, 
surface coating \cite{nesrullajev} and 
photoalignment \cite{ichimura,hwang,ho,kawatsuki} are only a few examples 
to mention.
When employing these techniques, topological defects are regarded as 
unwanted matters and thus they are eliminated artificially.
Under certain conditions, however, topological defects can have beneficial 
effects on the alignment control of liquid crystal molecules 
\cite{chiccoli1,berggren,chiccoli2,nelson1,vitelli1,santangelo1,fernandez,shin,xing,lopezleon} 
as well as of guest particles embedded in a liquid 
crystal medium \cite{ruan,musevic,yoon,skacej,guo,fukuda3,ravnik,bates}.
The positive effects result largely from topological constraints for the 
molecular orientational field ($=$ director field) that is confined in a 
closed curved substrate.
In fact, the director field over closed curved surfaces strongly depend 
on the positions and charges of topological defects distributed on the 
surfaces, which implies the controllability of the director field through 
the defect manipulation.
Such the utilities of topological defects may be enhanced when combined 
with the geometric potential effect.
It is known that topological defects in curved liquid crystal films 
experience a surface-curvature-induced potential, attractive or repulsive 
depending on the local geometry of the surface 
\cite{nelson2,seung,park,vitelli2,vitelli3,gomez}
\footnote{Similar curvature effects were discussed in the literature of 
quantum transport \cite{shima1,ono} and classical spin models \cite{baek} on 
curved surfaces.}.
Therefore, a better understanding of the correlation between curvature 
variation of the underlying surface and favourable configurations of 
topological defects will shed new light on molecular alignment control of 
liquid crystal films. 

\begin{figure}[t]
\begin{center}
\includegraphics[width=340pt]{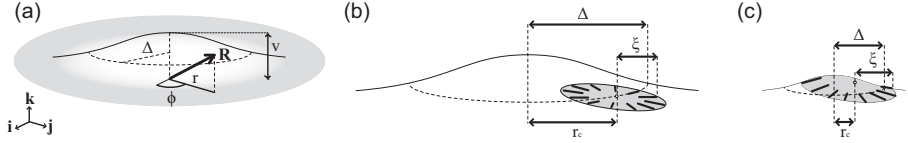}
\end{center}
\caption{\label{fig:geometry} (a) Sketch of a Gaussian bump characterized 
by the width $\Delta$ and the hight $v$.
The region inside (outside) the dashed circle with radius $\Delta$ possesses 
a positive (negative) Gaussian curvature.
(b, c) Disclination embedded in the Gaussian bump.
The coherence length $\xi$ of the nematic order fluctuation measures the 
spatial extent of desclinations (hatched regions) in which the director 
field $\bm n({\bm R})$ is arranged in a radial manner.}
\end{figure}

A Gaussian bump is one of the exemplary curved surfaces for studying the 
correlation mentioned above.
It has a continuous rotation symmetry around a vertical axis, showing a 
positive (negative) Gaussian curvature close to (far from) the top of the 
bump [see Fig. \ref{fig:geometry}(a)].
The coexistence of a positively curved region and a negatively one on the 
same surface allows a simultaneous observation of the curvature effects at 
the two distinct regions with oppositely-signed Gaussian curvature.
Earlier work suggested that when a nematic liquid crystal film is placed on 
a Gaussian bump, a point disclination with $+1$ charge is trapped to the 
top of the bump due to the curvature potential 
\cite{vitelli2,vitelli3}.
Contrariwise, we have found in the previous work that point disclinations 
are attracted to the inflection point of the bump regardless of their 
charges \cite{isaku}.
These two results appear controversial; a possible clue to resolve the 
controversy is the difference in the assumed bump widths between the two 
existing studies.
In the former study, the bump width is sufficiently small, so that the 
bump is covered almost entirely by a single disclination as depicted in 
Fig. \ref{fig:geometry}(c).
This situation is expressed by saying that the coherence length 
\cite{kleman1,wong}, $\xi$, of the nematic order fluctuation exceeds the 
scale of the bump width (see \S \ref{sec:method} for the definition of 
$\xi$).
In contrast, the latter study was based on the assumption that the bump 
width is larger enough than the coherence length and the system can involve 
many small disclinations [see Fig. \ref{fig:geometry}(b)].
It is thus reasonable to conjecture that the ratio of the bump width to the 
coherence length plays a decisive role in determining the stable position of 
disclinations on a Gaussian bump.
Furthermore, the two contrasting results imply the existence of a 
threshold ratio across which the stable position of disclinations alters.

In the present Letter, we prove our conjecture by showing a sudden shift 
in the stable position of a $+1$ disclination with increasing the bump width.
Once the bump width exceeds a threshold value, the stable position is 
suddenly switched from the top to the inflection point of the bump.
The positional shift is described by a power law function of the bump width, 
as is analogous to the second-order phase transitions of various physical 
systems whose properties are characterized by power-law behaviors.
The present result thus indicates a novel kind of phase transition that 
is peculiar to curved nematic films.

\section{Method}
\label{sec:method}
We consider point disclinations in a nematic liquid crystal membrane having a 
Gaussian bump geometry.
The bump is described by a position vector 
${\bm R(r,\phi)} = r \cos \phi \bm{i} + r \sin \phi \bm{j}
                     + v \exp (-r^2/2 \Delta^2) \bm{k}$,
where $\bm{i}$, $\bm{j}$, $\bm{k}$ are right-handed orthogonal basis vectors 
[see Fig. \ref{fig:geometry}(a)].
The constant $\Delta$, which we call the bump width, determines the 
inflection point of the bump at 
which $(\partial^2 {\bm R} / {\partial r}^2) \cdot {\bm k} = 0$.
A dimensionless parameter $\alpha \equiv v/\Delta$ quantifies the 
convexity of the bump; $\alpha = 0$ for the flat plane, and $\alpha \gg 1$ 
for a spiky bump.
We fix $\alpha \equiv 0.1$ throughout the paper, as a result of which 
spatial variation of the surface curvature is smooth and sufficiently 
slight to be accessible in experiments.

Our aim is to evaluate the stable position of a disclination constrained on 
the Gaussian bump.
It is solved by calculating the elastic free energy $F_{\rm d}$ of the 
director field ${\bm n}({\bm R})$ around the disclination core, since 
$F_{\rm d}$ takes a minimum at the stable position to be evaluated.
We assign the core of a $+1$ disclination at 
${\bm R}_{\rm c} = {\bm R(r_{\rm c},\phi_{\rm c})}$ with arbitrary 
$\phi_{\rm c}$, and assume that the disclination is endowed with the coherence 
length $\xi$ of the nematic order fluctuation.
The latter assumption ensures a radial pattern of $\bm n({\bm R})$ within 
a circular region $A$ having the radius $\xi$ centered at the core [i.e., 
the hatched regions depicted in Figs. \ref{fig:geometry}(b) and 
\ref{fig:geometry}(c)] \footnote{Precisely speaking, the area $A$ slightly 
deviates from an exact circle due to the finite curvature of the surface.
In fact, we define the perimeter of $A$ by the locus of points whose 
distances from the core are equal to $\xi$ in the sense of geodesic distance.
Therefore, the perimeter is distorted a bit from the exact circle.}.
The radial pattern of $\bm n({\bm R})$ is gradually distorted with increasing 
distance from the core; in fact, two-point correlation in the director 
orientation disappears if the points are well separated by a distance larger 
than $\xi$.
A typical value of $\xi$ is on the order of tens micrometers as can be 
estimated from polarization microscope images of liquid crystal films 
\cite{nehring1,shiwaku}.

The continuum elastic theory states that $F_{\rm d}$ associated with the 
disclination reads as \cite{santangelo2,kleman2}
\begin{eqnarray}
F_{\rm d} &=& h \int_A dA \left\{ \frac{K_1}{2} ({\bm D} \cdot {\bm n})^2
+ \frac{K_3}{2} ({\bm D} \times {\bm n})^2 + \frac{K_{24}}{2} 
{\bm D} \cdot [({\bm n} \cdot {\bm D}) {\bm n} -
{\bm n} ({\bm D} \cdot {\bm n})] \right\}
\mbox{,}
\label{eq:energy}
\end{eqnarray}
where $\bm D$ is a vector operator whose components $D_{\mu}$ are the 
covariant derivative \cite{shima2} on the curved surface.
The constant $h$ is the thickness of the liquid crystal film; $K_1$, 
$K_3$ and $K_{24}$ are the elastic constants associated with splay, bend 
and saddle-splay distortions of the field $\bm n$, respectively 
\cite{stewart}.
The integration range in Eq. (\ref{eq:energy}) is restricted to the finite 
circular region $A$, despite of slow decay in the elastic energy density with 
distance from the core; we will revisit this issue in \S \ref{sec:discussions}.
To avoid the divergence in $F_{\rm d}$ at the core, we introduce an inner 
cut-off $A_{\rm {core}}$ around the core according to the formula 
$A_{\rm {core}} = \pi K_1/2 k_{\rm B} (T_{\rm c} - T)$, in which 
$T_c$ is the nematic-isotropic transition temperature and $T$ is the 
temperature of the system \cite{cladis}.

In actual calculations, we set $\xi = 20 {\rm {\mu m}}$ 
and $h = 4 \  {\rm {\mu m}}$ by referring to the experimental 
observation \cite{wong,shah}.
The values of $K_1$ and $K_3$ are those of 
4-methoxybenzylidene-4$^{\prime}$-butylaniline (MBBA): $K_1 = 6$ pN, 
$K_3 = 7.5$ pN at room temperature \cite{stewart}.
Since the corresponding value of $K_{24}$ was not yet measured, we set 
$K_{24} = 4.9$ pN on the basis of the relation $K_{24} = (K_1 + K_2)/2$ 
that involves the twist coefficient $K_2 = 3.8$ pN for MBBA \cite{nehring2}.

\begin{figure*}[t]
\begin{center}
\includegraphics[width=330pt]{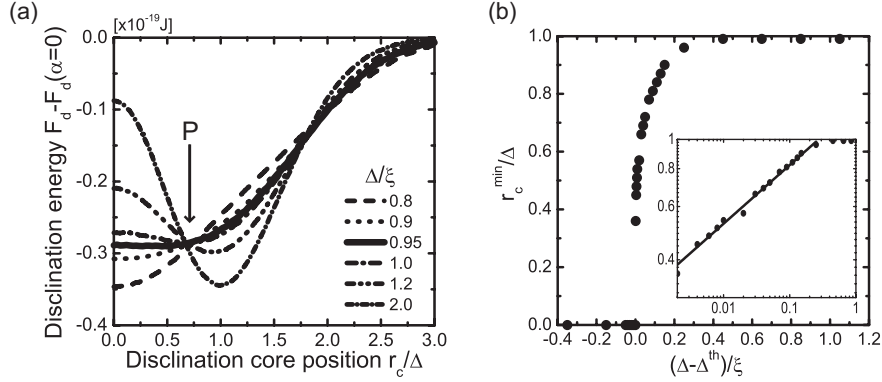}
\end{center}
\caption{\label{fig:energyplot} (a) Disclination energy $F_{\rm d}$ in 
deformed MBBA films as a function of the core position $r_{\rm c}$.
(b) Sudden shift in the stable position $r_{\rm c}^{\rm min}$ as a function 
of the bump width $\Delta$.
The threshold $\Delta^{\rm th}$ is estimated to be 
$\Delta^{\rm th}/\xi = 0.95$.
Inset: The same data in the log-log plot, showing a power law 
$r_{\rm c}^{\rm min} \propto \left| \Delta - \Delta^{\rm th} \right|^{\beta}$ 
with the exponent $\beta \sim 0.2$.}
\end{figure*}

\section{Results}
Figure \ref{fig:energyplot}(a) shows the disclination energy $F_{\rm d}$ 
as a function of the core position $r_{\rm c}$.
The normalized bump width, $\Delta/\xi$, ranges from $0.8$ to $2.0$.
For $\Delta/\xi = 2.0$, the curve of $F_{\rm d}$ is convex downward, showing a 
minimum at $r_{\rm c} \simeq \Delta$.
The energy scale of the potential depth corresponds to ${k_{\rm B} T}$ at a 
temperature of $T = 10^4 {\rm K}$, which is by far larger than the thermal 
excitation energy at room temperature.
Hence, this downward peak implies that a disclination is 
trapped to the annulus region having the radius of $r = \Delta$ 
surrounding the bump top [see Fig. \ref{fig:halocon}(a)].
We call this annular distribution of disclinations by the
\lq\lq halo\rq\rq\ phase, 
according to the discussion in Ref. \cite{isaku}.

What deserves attention is the $\Delta$-driven transition of the downward 
peak profile.
It follows from Fig. \ref{fig:energyplot}(a) that both the position and 
magnitude of the peak are insensitive to the decrease in $\Delta/\xi$ until 
it reaches a specific value of $\Delta/\xi = 0.95$.
A drastic change takes place at $\Delta/\xi = 0.95$; the downward peak in 
$F_{\rm d}$ is replaced by a plateau that extends from $r_{\rm c} = 0$ to 
$r_{\rm c} \sim 0.5 \Delta$.
This plateau means that the disclination can move freely within a finite 
circular region centered at the bump top.
It is also found that a further decrease in $\Delta/\xi$ gives rise to a new 
downward peak in $F_{\rm d}$ at $r_{\rm c} = 0$.
That is, when the bump width $\Delta$ is comparable to (or smaller than) the 
coherence length $\xi$, then the disclination is trapped at the bump top 
[see Fig. \ref{fig:halocon}(b)]; this result is in complete agreement with 
the earlier observation \cite{vitelli2,vitelli3}.
We call this defect configuration by the \lq\lq convergence\rq\rq\ phase.

We have seen that the peak of $F_{\rm d}$ appearing at 
$r_{\rm c} = r_{\rm c}^{\rm min}$ is 
gradually shifted from the inflection point 
$(r_{\rm c}^{\rm min} \sim \Delta)$ to the top 
($r_{\rm c}^{\rm min} = 0$) of the bump with increasing $\Delta$.
Figure 2(b) shows the $\Delta$-dependence of the normalized
peak position $r_{\rm c}^{\rm min}/\Delta$.
For a large bump width ($\Delta/\xi \gg 1$),
$F_{\rm d}$ takes the minimum at $r_{\rm c}^{\rm min}/\Delta \sim 1$
implying the halo phase.
The disclination halo shrinks with decreasing $\Delta$,
and finally it converges at the bump top when 
$\Delta = \Delta^{\rm th} \equiv 0.95 \xi$.
These results indicate the continuous phase transition
(i.e., of the second order)
of the stable disclination configuration on the Gaussian bump;
the system undergoes the transition from the halo phase
to the convergence phase across the threshold $\Delta^{\rm th}/\xi$.
This is the main finding of the present Letter.

It is interesting to note that the $\Delta$-dependence of 
$r_{\rm c}^{\rm min}$ in the halo phase is described by a power law 
of the form
\begin{equation}
r_{\rm c}^{\rm min} \propto \left| \Delta - \Delta^{\rm th} \right|^{\beta}.
\label{eq_shima}
\end{equation}
The inset of Fig.~2(b) evidences the power-law behavior of 
Eq.~(\ref{eq_shima})
with the exponent $\beta \sim 0.2$,
which holds over the two orders of the lateral axis.
We thus believe that the present result is a precursor of an
untouched configurational phase transition peculiar to curved liquid 
crystal systems, though further studies are needed to establish its 
universality and physical interpretation.

\section{Discussions}
\label{sec:discussions}
The transition of the stable defect configuration
we have demonstrated is a consequence of the following two
$F_{\rm d}$ variations caused by reducing $\Delta$: a slight 
$F_{\rm d}$-increase to the right of the arrow $P$ marked in Fig.~2(a), 
and a rapid $F_{\rm d}$-decrease to the left of $P$.
In this section, we give an outline of the physical mechanisms
of the two $F_{\rm d}$ variations, while a detailed formulation
that describes them will be given elsewhere.

The rapid decrease in $F_{\rm d}$ to the left of $P$ originates from 
a suppression of the splay deformation of the director field 
${\bm n} (\bm{R})$ around the bump top.
Suppose that a $+1$ disclination is placed near the top
of a broad bump as depicted in Fig.~\ref{fig:geometry}(c).
Since $\xi$ and $\alpha = v/\Delta$ are constants, a reduction in $\Delta$ 
leads to a lowering of the bump hight $v$ remaining the disclination area 
$A$ be fixed.
As a result, the disclination is allowed to cover the whole bump.
In the latter situation, adjacent directors (thick line segments in the 
plots) distributed close to the outer edge of $A$ tend to be parallel to 
each other.
Eventually, the first integrated term in Eq. (\ref{eq:energy}), 
$\propto (\bm{D}\cdot\bm{n})^2$,
which represents the contribution from the splay deformation of $\bm{n}$,
is suppressed by reducing $\Delta$.
This is the reason why $F_{\rm d}$ at $r_{\rm c}=0$ drops off when 
$\Delta$ decreases.

A parallel argument to the above accounts for the slight increase
in $F_{\rm d}$ to the right of $P$ in Fig.~\ref{fig:energyplot}(a).
Suppose a $+1$ disclination on the skirt (i.e., at $r_{\rm c} \sim \Delta$)
of the bump.
If the disclination area $A$ intersects the circle $r=\Delta$ around the 
bump top,
then it covers a positively-curved region ($r<\Delta$)
and a negatively-curved one ($r>\Delta$).
At the negatively curved region, a reduction in $\Delta$ yields an 
enhancement of the splay deformation which leads to the slight increase 
in $F_{\rm d}$ at $r\sim \Delta$.
A mathematical proof of the splay enhancement has been given in 
Ref. \cite{isaku}.

The data shown in Figs. \ref{fig:energyplot}(a) and \ref{fig:energyplot}(b) 
are derived from a specific Gaussian bump shape characterized by the parameter 
$\alpha = v/\Delta \equiv 0.1$ (see Fig. \ref{fig:geometry}(a)).
We also confirmed that, when $0.05 < \alpha < 0.5$, the system exhibits the 
same configurational transition across $\Delta^{\rm th}$.
A slight increase in $\Delta^{\rm th}$ with $\alpha$ was observed within the range 
of $\alpha$, reflecting further splay suppressions and enhancements in 
disclinations whose cores locate at $r_{\rm c} = 0$ and $r_{\rm c} = \Delta$, 
respectively.
In contrast, the value of $\beta$ is insensitive to the change in $\alpha$.

It should be stated that our discussion is based on the model given by 
Eq. (\ref{eq:energy}), in which contributions from the outer region than the 
circle $A$ are ignored.
In actual nematic layers, a $+1$ disclination has a profile of the elastic 
energy density that decays logarithmically with distance from the core.
Hence, contributions far from the core may be relevant to the stable position 
of the disclination, thus should be taken in account in order to obtain a 
quantitatively accurate value of $\Delta^{\rm th}$.
The geometric potential approach developed in Ref. \cite{vitelli4} will be a 
clue to solve this issue.

\begin{figure*}[t]
\begin{center}
\includegraphics[width=350pt]{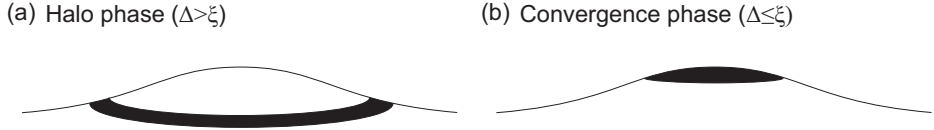}
\end{center}
\caption{\label{fig:halocon} Distribution patterns of disclinations in the 
halo phase (a) and convergence phase (b).
In either phase, disclinations are trapped in the dark region.}
\end{figure*}

Before closing the Letter, we suggest the following system could allow 
experimental tests of our theoretical predictions; it is a thin liquid crystal 
film coating a randomly corrugated substrate.
The film involves many disclinations and bumps with various sizes; therefore, 
we will observe simultaneously the two effects, halo and convergence patterns 
depicted in Fig. \ref{fig:halocon}, by tuning the surface corrugation.
We also mention that such the corrugated substrate can be synthesized by polymer 
adhesion with patterned surfaces as demonstrated in Ref. \cite{chan}.

\section{Conclusion}
We have found a continuous transition of defect configurations in nematic 
films confined on a Gaussian bump.
Depending on the ratio $\Delta/\xi$, the system exhibits either of two 
defect distribution patterns: the halo phase ($\Delta/\xi>1$) and 
convergence phase ($\Delta/\xi \le 1$).
The stable position $r_{\rm c}^{\rm min}$ of defects turned out to obey the 
power law 
$r_{\rm c}^{\rm min} \propto \left| \Delta - \Delta^{\rm th} \right|^{\beta}$ 
with the exponent $\beta \sim 0.2$ under the present condition.
Our findings suggest the possibility of defect manipulation via artificial 
deformation of liquid crystal films.

\section*{Acknowledgments}
We would like to thank K.~Yakubo, H. Orihara, J. Fukuda and Y. H. Na for 
fruitful discussions.
IH is thankful for the financial support from the Japan Society for the 
Promotion of Science for Young scientists.
HS acknowledges M. Arroyo for his hospitality during the stay in UPC.
This work is supported by the Kazima foundation and a Grant-in-Aid for 
Scientific Research from the Japan Ministry of Education, Science, Sports and 
Culture.

\end{document}